\documentclass[%
 reprint,
 amsmath,amssymb,
 aps,
]{revtex4-2}

\usepackage{graphicx}
\usepackage{dcolumn}
\usepackage{bm}
\usepackage{xcolor}
\usepackage{soul}
\usepackage{threeparttable}

\begin{document}

\preprint{APS/123-QED}

\title{Hong-Ou-Mandel two-photon x-ray states}

\author{Liam T. Powers}
\author{Stephen M. Durbin}%
 \email{durbin@purdue.edu}
\affiliation{%
 Department of Physics \& Astronomy, Purdue University, West Lafayette IN 47907-2036
}%

\date{\today}

\begin{abstract}
We have observed Hong-Ou-Mandel interference of high brightness synchrotron x-rays with a  Mach-Zehnder interferometer, yielding two-photon states of potential interest for x-ray quantum optics.
\end{abstract}

\maketitle

The essence of single-photon states is perhaps best exemplified by Dirac’s famous claim about Michelson interferometry that “Each photon then interferes only with itself. Interference with different photons can never occur”\cite{dirac_principles_1930}. Parametric down-conversion (PDC) has since enabled the reliable generation of pairs of photons and interference involving two photons was reported  by Hong, Ou, and Mandel (HOM) \cite{hong_measurement_1987}. Two indistinguishable photons at separate entrance ports of a beam splitter generate two-photon interference, with destructive interference preventing the two photons from taking different exit paths (Fig 1a). The Dirac claim was then updated to clarify that two-photon (or multi-photon) states only interfere with themselves \cite{kim_two-photon_2003,noauthor_multi-photon_2007}. Multi-photon states are now part of various quantum applications, including teleportation, optical computing, and quantum metrology \cite{bouwmeester_experimental_1997,kok_linear_2007,boto_quantum_2000}.

\begin{figure}[b]
\includegraphics[width=1\linewidth]{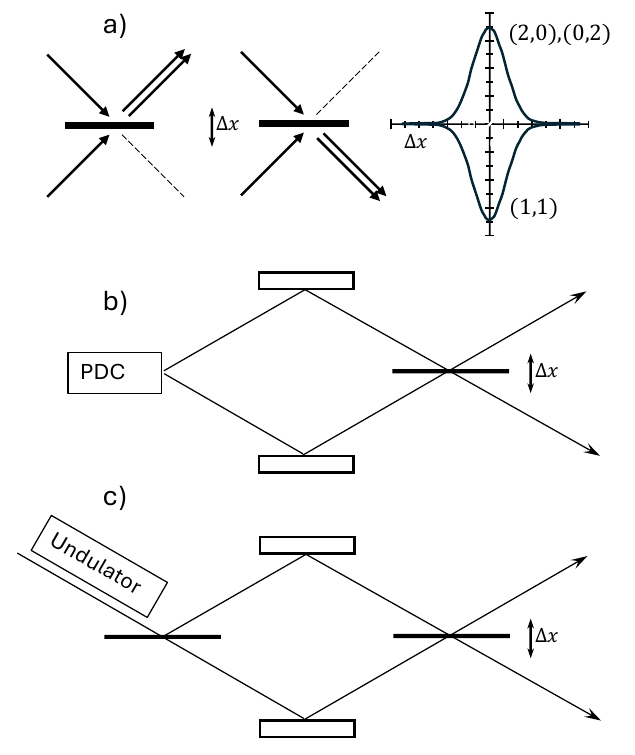}
\caption{\label{fig:1} Hong--Ou--Mandel (HOM) geometries. 
a) A 50-50 beam splitter with indistinguishable incident photons produces either two photons exiting from one output port $(2,0)$ or two photons exiting from the other $(0,2)$. The plot shows how the $(2,0)$ and $(0,2)$ rates compare to the $(1,1)$ rate as the beam splitter is displaced ($\Delta x$).
b) Schematic HOM layout utilizing parametric down-conversion (PDC) and two mirrors to direct photons onto a beam splitter.
c) Synchrotron layout utilizing a Mach-Zehnder interferometer, where the undulator radiation is split into two beams that are redirected by Bragg mirrors onto the HOM beam splitter.
}
\end{figure}

Analogous progress with x-ray photons has been quite limited. Nearly all x-ray science is still in the Dirac single-photon-interference phase despite the enormous advances at synchrotron and free electron laser sources \cite{stohr_overcoming_2019}.  The exceptions include Hanbury Brown - Twiss observations and x-ray parametric down-conversion \cite{freund_parametric_1969,hartley_x-ray_2025,gluskin_classical_1999, gorobtsov_diffraction_2018, singer_hanbury_2013,  singer_intensity_2014}. There has even been a proposal to employ HOM interference to determine the attosecond pulse duration of PDC-generated x-rays\cite{volkovich_PDC_HOM_2020}. Such PDC measurements are technically possible, but are challenging due to experimental count-rates typically below 1 Hz.

We report here the observation of x-ray two-photon interference using a Mach-Zehnder interferometer. Standard HOM interference of PDC-generated photons (Fig 1b) yields maximally entangled N00N quantum states \cite{dowling_quantum_2008}, which would be of great interest for quantum x-ray applications. The characteristic HOM dip, however, can also be seen with thermal or chaotic light, and even with classical light \cite{Liu2014ThermalHOM, Chen2011ShihThermal, Chen2013TrueThermal}. After presenting the experimental results below, we will argue that radiation from an undulator is best described as chaotic light that can yield quantum two-photon interference.

Two-photon interference with a synchrotron source has these requirements: the source must generate indistinguishable photons, these pairs are split to take separate paths, a scanning beamsplitter varies their spatial or temporal overlap to sample the region where their output trajectories are indistinguishable, and a multi-photon detection system counts the number of photons in each exit beam for each synchrotron pulse. The need to split and recombine the beams naturally calls for a Mach-Zehnder interferometer design, as illustrated in Fig 1c.

HOM interference requires two photons in the same mode, since otherwise they could be distinguishable. The mean number of photons per mode with wavelength $\lambda$ for a synchrotron source is a function of the peak brightness $B$:

\begin{equation}
n_{\mathrm{ph}} = \frac{B_{peak}\,\lambda^{3}}{4\sqrt{2}\,c}
\label{eq1}
\end{equation}
where c is the speed of light  and $n_{\mathrm{ph}}$ is known as the degeneracy factor \cite{stohr_two-photon_2017}. At high-brightness synchrotrons such as the APS-U at Argonne National Laboratory, where this work was conducted \cite{shi_measurements_2025}, the degeneracy factor can now be greater than unity for certain energies. Assuming a thermal distribution the probability of k photons being found in a given mode is \cite{loudon_quantum_2000}

\begin{equation}
P_k = \frac{n_{\mathrm{ph}}^{\,k}}{\left(1 + n_{\mathrm{ph}}\right)^{k+1}}
\label{eq2}
\end{equation}
For example, at $n_{ph}=2$ more than 22$\%$ of photons are born in a doubly occupied mode, but there is a large background from singly-occupied and other modes. Two-photon interference measurements thus require good statistics to separate the signal from background.

To observe interference, the synchrotron beam is set to the energy $E=8$ keV with bandwidth $\Delta E = 74$ meV utilizing the upstream beamline monochromators; see Supplemental Materials for details. The specially constructed Mach-Zehnder interferometer had a single monolith for the two Bragg x-ray mirrors and thinned silicon crystals for the two beam splitters, all oriented for Si (400) diffraction \cite{powers_variable_2025}. The Bragg geometry was chosen to minimize beam absorption losses.

\begin{figure}[t]
\includegraphics[width=0.8\linewidth]{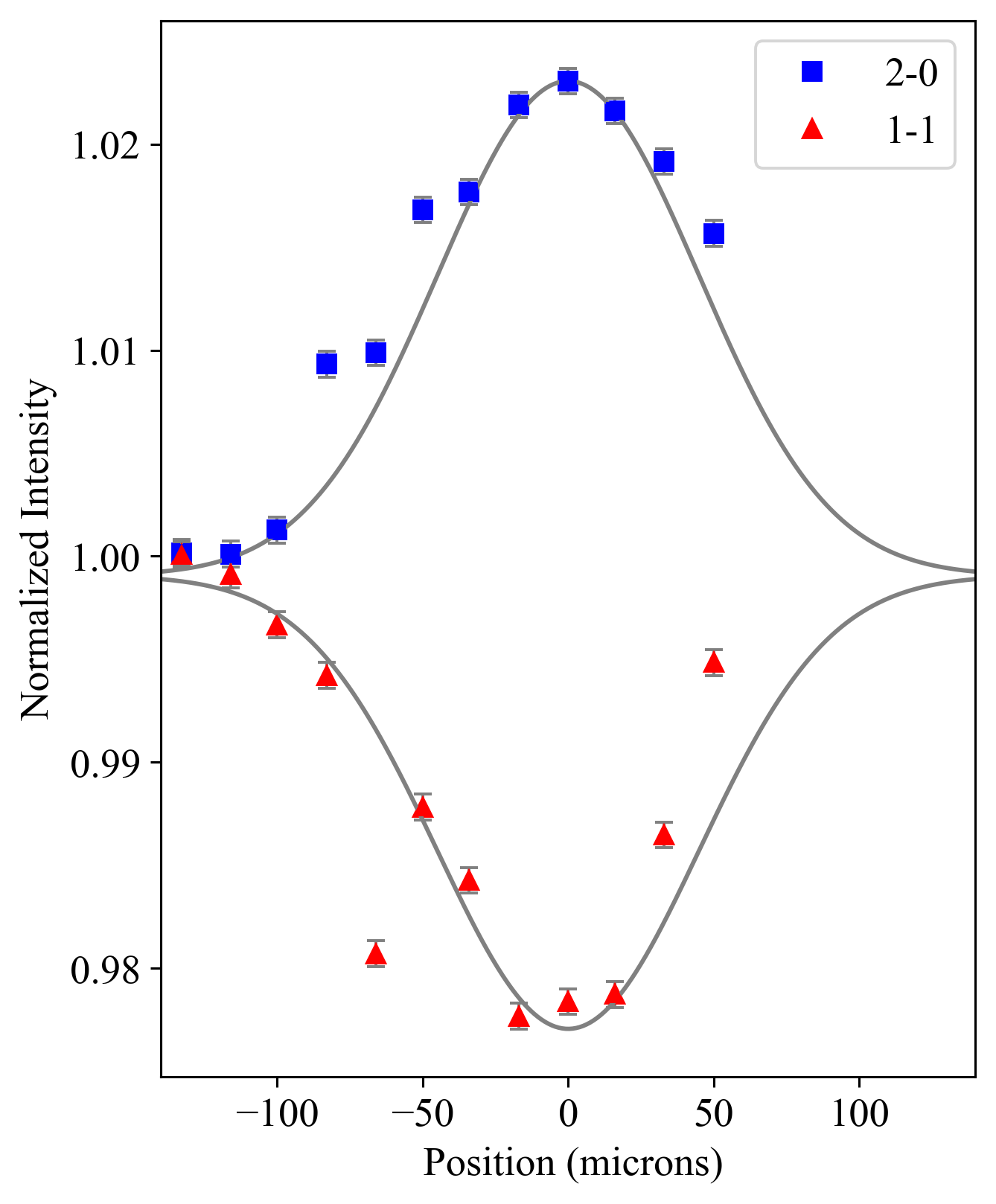}

\caption{\label{fig:4} 
 HOM x-ray interference. The $(1,1)$ and the $(2,0)$ $\&$ $(0,2)$ rates are plotted versus beam splitter position $\Delta x$, revealing the HOM dip and peak. The data are fit by a Gaussian $\exp[{-\Delta x^2 /l_c^2}]$ where $l_c$ is the lateral coherence length as defined by upstream slits \cite{Lee2006SpatialLabeling}. The best fit shown used a FWHM of $106$ $\mu m$, in good agreement with the 100 $\mu m$ slit setting.
}
\end{figure}

Both splitters are controlled by precision linear and rotary stages, critical for ensuring that all elements are diffracting optimally and the two equal-length beam paths intersect at the second beam splitter. Following the original HOM protocol  (Fig \ref{fig:1}), the beam splitter is translated by fixed values of $\Delta x$ to vary its overlap with the two beams. Beam slits were adjusted so that on average only one photon was in each beam striking the final beam splitter for each synchrotron pulse, a five-order-of-magnitude reduction from the full undulator beam intensity. A typical beam size was 100 $\mu m$ by 100 $\mu m$, less than the 140 $\mu m$ lateral coherence length of the source.

\begin{figure}[t]
\includegraphics[width=0.8\linewidth]{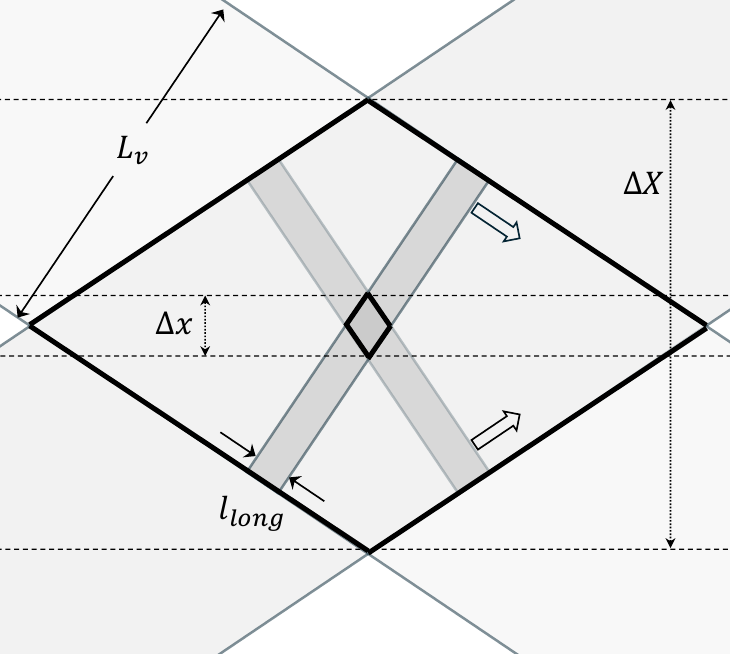}

\caption{\label{fig:5} 
Schematic of x-ray beam overlap. The large diamond maps out the region where the two incident beams overlap in space; the beam width $L_v$ is set by the upstream vertical slit. A beam width of 100 $\mu m$ corresponds to a geometric height $\Delta X$ of $\sim 120$ $\mu m$. The smaller diamond is the instantaneous region of overlap of two simultaneous x-rays, as defined by the 8 $\mu m$ coherence time. $\Delta x$ is the beam splitter range that samples temporally overlapping photons, with a geometric height of $\sim 14$ $\mu m$. $\Delta X$ spans the entire overlap region including where the photons cannot be simultaneously present, which closely matches the region where interference was observed.
}
\end{figure}

An ideal interference measurement requires counting the integer number of photons at both detectors from each synchrotron bunch. The voltage outputs of photon-number-resolving avalanche photodiode (APD) detectors are digitized and analyzed before the next pulse arrives \cite{powers_counting_2025}. This permitted the independent analysis of 13 million discrete events per second ($i.e.$ the synchrotron bunch frequency), a new experimental modality where each synchrotron  bunch yields a unique measurement. Output signals are converted to an integer number of photons (0, 1, 2...) based on prior calibration, so that one synchrotron bunch yields one pair of integers. 

In principle an HOM experiment simply requires counting the number of (1,1), (2,0), and (0,2) pairs recorded in a given time interval as a function of the beam splitter displacement. Figure 2 is a plot of such a measurement, which reveals a dip in the normalized  (1,1) coincidence counts versus $\Delta x$. Also plotted is the corresponding peak in the combined (2,0) and (0,2) counts, clearly showing that the (1,1) output is being converted into (2,0),(0,2) outputs. This is consistent with the HOM explanation that the (1,1) output is suppressed for indistinguishable photons. Each data point is the sum of three scans each of 2 seconds duration, with error bars determined from counting statistics. Larger fluctuations are caused by transient variations in the diffracting elements; see Supplementary Materials for more detail. With the incident flux set to one photon per beam per bunch, this dip corresponds to the interference  of $\sim3 \times 10^{4}$ photon pairs per second. Having a dip of only $\sim2\%$ means that most pairs of photons reaching the final beam splitter are not in the same mode. Much of this is due to known issues with the beamline optics and the storage ring that limit the effective brightness and hence the photon degeneracy.

\begin{figure}[t]
\includegraphics[width=0.8\linewidth]{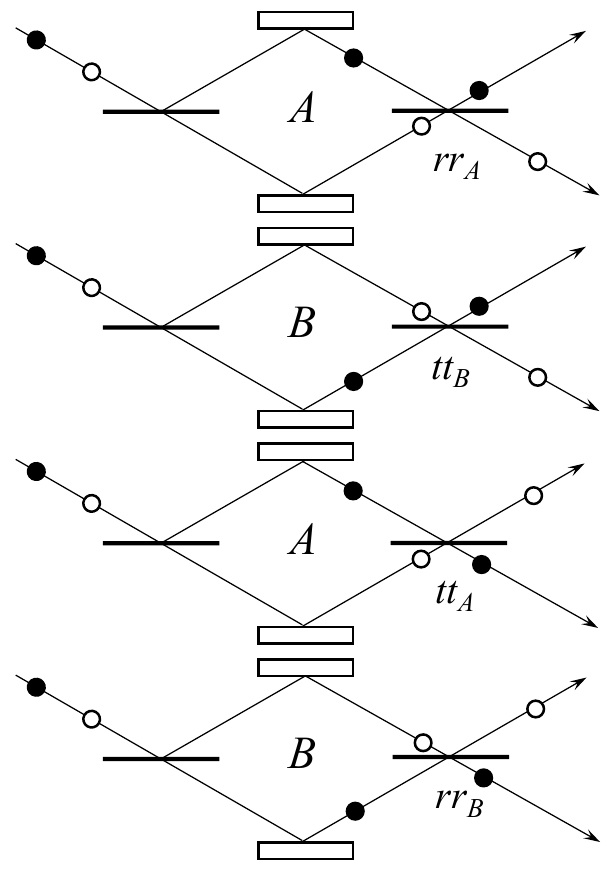}

\caption{\label{fig:5} 
Indistinguishable output trajectories for time separated pairs. Two separated photons (filled and empty circles) enter the Mach-Zehnder interferometer and can take different paths. Case A is when the leading photon takes the lower path, and for Case B it takes the upper path. These are also shown approaching the HOM beam splitter, and again on their exit paths towards the detectors. Note that Case A with both reflected ($rr_A$) is indistinguishable from Case B with both transmitted ($tt_B$), and the same for $tt_A$ and $rr_B$.
}
\end{figure}

A sketch of the real-space overlap of the two incident beams at the final beam splitter is shown in Figure 3; the beam splitter would be represented by a horizontal line in this image. Indistinguishability requires that the two photons must at least partially overlap the same region on the beam splitter, as outlined by the large diamond of height $\Delta X$. If the two photons arrive simultaneously, however, the overlap of their longitudinal coherence lengths ($\sim8$ $\mu m$, or $\sim30 $ fs) is limited by the smaller diamond of height $\Delta x$. Comparison with the data (Fig 2) reveals that the observed width of the HOM dip is determined by the entire overlap region (the large diamond) and cannot be explained by the interference of simultaneous photons. That is, this  version of HOM interference is due to photons separated by more than their coherence times, potentially as much as the $\sim200$ ps synchrotron bunch duration.

When Hong  \textit{et al.} observed their dip, they utilized pairs of photons each in singly occupied number states generated by PDC, the photon pairs arrived simultaneously at the beam splitter, and after exiting reached their detectors at the same time. Later it was shown that the photons do not have to be born at the same time or even from the same source \cite{RARITY.J.G.1995, ZUKOWSKI_1995_entangling_photons, kaltenbaek_experimental_2006}, and they do not have to arrive at the beam splitter or detectors simultaneously \cite{kim_two-photon_2003, pittman_can_1996, kim_hongoumandel_2020}. What is necessary is that the trajectories when both photons are transmitted (\textit{tt}) and when both photons are reflected (\textit{rr}) are indistinguishable. Then the probability amplitudes for $tt$ and $rr$ cancel, leaving only the doubly occupied $tr$ and $rt$ exit beam states. We assume this mechanism also explains the observed x-ray dip.

Figure 4 illustrates this process for the Mach-Zehnder interferometer. Consider two photons from the same bunch and in the same mode but with a time delay $\Delta t$. For those cases where the two photons take different paths, the lead photon could take the lower path (Case A) or the upper path (Case B). Looking only at the exit trajectories after the final beamsplitter, it is seen that $rr_A$ is indistinguishable from $tt_B$, as are $rr_B$ and $tt_A$. That is, not knowing which path is taken in the interferometer creates the HOM interference. 

The unresolved question is, do these exit states form something like a coherent N00N state, or are they in some other type of state or mixture of states? This question is important because a N00N state is a superposition that could be a source of entangled x-rays for probing quantum materials. On the other hand, it is possible to observe a reduced HOM dip with classical light, which would not engender entanglement. Resolving this question will ultimately require tests of the output beams for entanglement, such as an x-ray adaptation of Bell inequality tests \cite{clauser_proposed_1969}.

It is worth considering how these x-rays are generated when an electron bunch transits an undulator at a synchrotron source, where radiation occurs stochastically from uncorrelated electrons. This rules out x-rays being in a coherent state, which would require an x-ray free electron laser source. Undulator radiation is described as “pseudo-thermal” \cite{Howells_1993}. Two-photon interference observations have been reported with visible light pseudo-thermal sources. While thermal light always includes a background of photons that do not interfere \cite{kwiat_correlated_1990}, those that do exhibit behavior similar to PDC entangled photons. In particular, the interference pattern from slits varies with $\lambda/2$ instead of $\lambda$ \cite{ boto_quantum_2000, kim_spatial_2006,dangelo_two-photon_2001, dopfer_brillouin_1995}. This has also been demonstrated with pseudo-thermal sources \cite{scarcelli_two-photon_2004}.  Furthermore, this suggests that x-ray two-photon states could diffract from a crystal with an effective wavelength of $\lambda /2$, providing a straight-forward means to isolate the two-photon states from the large background \cite{durbin_proposal_2022, stohr_overcoming_2019}.

The x-ray data arise from bunches where only two photons reach the detectors. While it is possible for a single electron to produce both x-rays, those would be simultaneous which is inconsistent with the data. Hence, they must arise from two different electrons. Since photons must be in the same mode for two-photon interference, the two electrons effectively follow the same path through the undulator. Simple analysis considering the electron density in the bunch (see Supplemental Materials for beam specifications) shows that a typical linear path of cross-sectional area $\lambda^2$  through the length of the bunch is likely to be occupied by more than one electron. Thus, we can expect multiple indistinguishable photons from typical bunches. 

Classical derivations of multi-photon interference consider a definite relative phase between the electric field amplitudes of the two waves \cite{mandel_photon_1983}. This could be an issue here if the two photons were generated by the same electron, but since they in fact arise from two time-separated uncorrelated electrons it is unlikely for a definite relative phase to be measurable. We therefore consider a classical wave explanation to be unlikely.

Advances in x-ray sources have led to a growing interest in multi-photon diffraction processes. This includes experimental progress at an x-ray free electron laser (XFEL) \cite{wu_stohr_2016} and theoretical considerations of stimulated multi-photon diffraction at high intensities \cite{stohr_two-photon_2017,stohr_overcoming_2019}. This complements extensive progress made with PDC-generated x-rays \cite{shwartz_PRL_2012,hartley_x-ray_2025} in recent years. The two-photon x-ray states created by HOM interference offers a new type of multi-photon state. It requires high brightness to achieve sufficient photon degeneracy, but not the high instantaneous intensities of an XFEL. The yields are substantially greater than what is currently achievable with x-ray PDC, but background issues are more challenging. The potential areas of application are likely to be different than with XFEL and PDC multi-photons. Future efforts would include increasing the HOM dip with improved x-ray optics, exploring the unique absorption properties of two-photon states \cite{roger_N00N_2016}, determining the degree of entanglement after devising appropriate Bell tests, and ultimately utilizing entangled x-ray probes to characterize entangled quantum materials \cite{irfan_neutron_2021}.

\begin{acknowledgments}
The experiments at APS-U Sector 27 (Argonne) benefited greatly from the efforts of Diego Casa, Jung Ho Kim, and Thomas Gog. We also acknowledge valuable contributions at Purdue from Aidan Jacobsen and Madeline Kwasniewski. This research was supported by the U.S. Department of Energy through Contract No.~DE-SC0023176 and was performed using beam time awarded at the Advanced Photon Source under DOI~\href{https://doi.org/10.46936/APS-191314/60015045}{10.46936/APS-191314/60015045}. The Advanced Photon Source is a U.S.~Department of Energy (DOE) Office of Science user facility operated for the DOE Office of Science by Argonne National Laboratory under Contract No.~DE-AC02-06CH11357.
\end{acknowledgments}

\bibliography{x_HOM2}
\end{document}